\documentclass[10pt,conference]{IEEEtran}
\IEEEoverridecommandlockouts
\usepackage{dirtytalk}
\usepackage{threeparttable}
\usepackage{amsmath,graphicx,amssymb,bm}
\usepackage{multirow}
\usepackage{booktabs}
\usepackage{xcolor}
\usepackage{colortbl}
\usepackage{multirow}
\usepackage{units}
\definecolor{royalblue}{rgb}{0.26, 0.41, 1}
\definecolor{black}{rgb}{0, 0, 0}
\definecolor{cobalt}{rgb}{0, 0, 0}
\definecolor{darkblue}{rgb}{0, 0, 0}

\definecolor{lgray}{RGB}{211, 211, 211}
\definecolor{dimgray}{RGB}{105, 105, 105}
\definecolor{whitesmoke}{RGB}{245, 245, 245}

\definecolor{modMRCcolor}{rgb}{0.9, 0.5, 0.9}
\definecolor{modMRCcolor2}{rgb}{0.9, 0.4, 0.1}
\definecolor{ZFCcolor}{rgb}{0.5, 0.5, 0.5}

\title{Massive MIMO Channel-aware Decision Fusion Aided by Reconfigurable Intelligent Surfaces}
\author{
\IEEEauthorblockN{
Domenico Ciuonzo\IEEEauthorrefmark{1},
Alessio Zappone\IEEEauthorrefmark{2},
Marco Di Renzo\IEEEauthorrefmark{3},
Linlong Wu\IEEEauthorrefmark{4}
}
\IEEEauthorblockA{
\IEEEauthorrefmark{1}\textit{DIETI, University of Naples Federico II, Italy},\\
\IEEEauthorrefmark{2}\textit{DIEI ``Maurizio Scarano'', University of Cassino and Southern Lazio, Italy \& CNIT, Italy},\\
\IEEEauthorrefmark{3}\textit{Laboratoire des Signaux et Syst\`emes, CentraleSup\'elec, CNRS, Universit\'e Paris-Saclay, France}\\
\IEEEauthorrefmark{4}\textit{SnT, University of Luxembourg, Luxembourg}
}

\thanks{The research leading to these results has received funding from Project ``Garden'', CUP H53D23000480001 funded by EU in NextGeneration EU plan through the Italian ``Bando Prin 2022 - D.D. 104 del 02-02-2022`` by MUR.
This manuscript reflects only the authors’ views and opinions and the Ministry cannot be considered responsible for them.
M. Di Renzo was partly supported by the European Commission through the Horizon Europe projects COVER (grant agreement no. 101086228), UNITE (grant agreement no. 101129618), and INSTINCT (grant agreement no. 101139161), as well as by the Agence Nationale de la Recherche (ANR) through the France 2030 project ANR-PEPR Networks of the Future (grant agreement NF-YACARI 22-PEFT-0005), and by the CHIST-ERA project PASSIONATE (grant agreements CHIST-ERA-22-WAI-04 and ANR-23-CHR4-0003-01). L. Wu acknowledges the Luxembourg National Research Fund (FNR) through the BRIDGE project S3 under grant C22/IS/17412681/S3.
}
}

\begin{document}
%
\maketitle
\begin{abstract}
This paper investigates channel-aware decision fusion empowered by massive MIMO systems and reconfigurable intelligent surfaces (RIS). By integrating both, we aim to improve goal-oriented (fusion) performance despite the unique propagation challenges introduced. Specifically, we investigate traditional favorable propagation properties in the context of RIS-aided Massive MIMO decision fusion. The above analysis is then leveraged ($i$) to design three sub-optimal simple fusion rules suited for the large-array regime and ($ii$) to devise an optimization criterion for RIS reflection coefficients based on long-term channel statistics. Simulation results confirm the appeal of the presented design.
\end{abstract}
\section{Introduction}
\label{sec:intro}

The Internet of Things (IoT) envisions the extensive integration of small, sensor-equipped devices that possess both processing and communication capabilities, transforming their role in everyday applications~\cite{Ciuonzo2019book}. This technology is poised to revolutionize the wireless communications \& sensing industries. At the core of IoT systems are Wireless Sensor Networks (WSNs), which function as the primary sensing framework. A key research area in this domain is Distributed Detection (DD), an established topic with a broad range of applications, from cognitive radio~\cite{Chen2019} to industrial setups~\cite{tabella2024bayesian}.

Recently, considerable research has focused on enhancing energy efficiency and simplifying processing by integrating intelligent reflective surfaces with \emph{massive MIMO} technology, referred to here as RISmMIMO. While prior studies have explored various aspects of RISmMIMO systems—such as performance with large arrays~\cite{wang2021wireless}, asymptotic behavior as both the number of antennas and RIS elements increase~\cite{wang2023massive}, or when only the RIS elements scale~\cite{wang2023massive}—the potential synergy between RIS and massive MIMO in the context of DD remains largely underexplored.

In contrast, studies on decentralized inference have examined the positive impact of large arrays on end-to-end inference performance, yielding similar benefits to those observed in cellular multi-user systems~\cite{ciuonzo2015,shirazinia2016massive,dey2020wideband}. Recent research has explored integrating RISs into WSNs for tasks such as DD~\cite{mudkey2022wireless}, distributed estimation ~\cite{Ahmed2022}, and nomographic function computation~\cite{fang2021}. However, these works focus on ($i$) non-asymptotic array setups and ($ii$) RIS designs based on instantaneous channel state information (CSI).

Hence, the \emph{main contributions} of this work are as follows.
We study DD with sensors transmitting decisions to a Fusion Center (FC) over a multiple-access channel, where the FC is equipped with a large receive array, forming a distributed massive MIMO setup~\cite{ciuonzo2015}. Information alignment is facilitated by a suitably-designed RIS~\cite{huang2019,direnzo2020}. 
Unlike traditional DD mMIMO setups, where linear rules like MRC are effective, this is not the case for RISmMIMO. Hence, we propose \emph{two key contributions}: ($i$) three alternative fusion rules that leverage the benefits of massive arrays at the FC with RIS; ($ii$) a RIS design based on long-term channel statistics using a majorization-minimization approach~\cite{sun2016majorization}, beneficial across all fusion rules.
Both contributions align to a \emph{two-timescale} design philosophy~\cite{pan2022overview}.

The paper is organized as follows. Sec.~\ref{sec:system_model} presents the system model, followed by the fusion rule design for the RISmMIMO scenario in Sec.~\ref{sec_fusion_rule_RIS}. Sec.~\ref{sec_RIS_design_LT} outlines the proposed long-term RIS design, with simulation results provided in Sec.~\ref{sec:sim_results}. Finally, Sec.~\ref{Conclusions} concludes with future research directions.%
\footnote{\textbf{Notation} -- Bold lower-case letters represent vectors, while bold upper-case letters represent matrices. The operators $\mathbb{E}\{\cdot\}$,
$\mathrm{var\{\cdot\}}$, $\mathrm{Cov(\cdot)}$, $(\cdot)^{T}$, $(\cdot)^{\dagger}$, $\Re\left(\cdot\right)$, $\angle(\cdot)$ and $\left\Vert \cdot\right\Vert $ denote expectation, variance, covariance, transpose, conjugate transpose, real part, phase, and Euclidean norm, respectively.
The symbols $\bm{O}_{N\times K}$ and $\bm{I}_{N}$ refer to the $N\times K$ null matrix and the $N\times N$ identity matrix, respectively.
Similarly, $\bm{0}_{N}$ and $\bm{1}_{N}$ represent the null vector and ones vector of length $N$. $\mathrm{diag}(\bm{a})$ refers to a diagonal matrix with $\bm{a}$ on the main diagonal.
$\Pr(\cdot)$ and $p(\cdot)$ are used for probability mass functions (pmfs) and probability density functions (pdfs), while $\Pr(\cdot|\cdot)$ and $p(\cdot|\cdot)$ indicate their conditional forms. 
The notation $\mathcal{N}_{\mathbb{C}}(\bm{\mu},\bm{\Sigma})$ represents a proper complex normal distribution with mean vector $\bm{\mu}$ and covariance matrix $\bm{\Sigma}$, and the symbol $\sim$ indicates \say{distributed as}.}

\section{System Model}\label{sec:system_model}

We consider a distributed binary hypothesis test with $K$  sensors to distinguish between hypotheses $\mathcal{H}\triangleq\{\mathcal{H}_{0},\mathcal{H}_{1}\}$ (e.g., absence/presence of a phenomenon). 
Each sensor $k\in\mathcal{K}\triangleq\{1,2,\ldots,K\}$ makes a local binary decision $\xi_{k}\in\mathcal{H}$ based on its observations, without assuming mutual independence given Hi. 
The $k$th sensor, $k\in\mathcal{K}\triangleq\{1,2,\ldots,K\}$, takes a binary local decision $\xi_{k}\in\mathcal{H}$ about the observed phenomenon on the basis of its own measurements. Here we do not make any conditional (given $\mathcal{H}_{i}\in\mathcal{H}$) mutual independence assumption on $\xi_{k}$. 
Each decision $\xi_{k}$ is mapped into $x_{k}\in{\cal X}=\{-1,+1\}$, using Binary Phase-Shift Keying (BPSK): without loss of generality we assume that $b_{k}=\mathcal{H}_{i}$
maps into $x_{k}=(2i-1)$, $i\in\{0,1\}$.
The quality of the WSN is described by the conditional joint pmfs $P(\bm{x}|\mathcal{H}_{i})$, and we define $P_{D,k}\triangleq P\left(x_{k}=1|\mathcal{H}_{1}\right)$ and $P_{F,k}\triangleq P\left(x_{k}=1|\mathcal{H}_{0}\right)$ as the detection and false alarm probabilities, respectively, with $P_{D,k}\geq P_{F,k}$.

The sensors communicate with a Fusion Center (FC) with $N$ antennas via a wireless flat-fading multiple access channel~\cite{Ciuonzo2012}, assisted by an RIS with $M$ elements. 
We consider a large-array setup where $N \gg K$.
Let $\bm{H}^{d}\in\mathbb{C}^{N\times K}$ ($\bm{h}_{k}^{d}\in\mathbb{C}^{N}$ being the $k$th sensor contribution), $\bm{H}^{r}\in\mathbb{C}^{M\times K}$ ($\bm{h}_{k}^{r}\in\mathbb{C}^{M}$ being the $k$th sensor contribution) and $\bm{G}\in\mathbb{C}^{N\times M}$ be the equivalent channels from the WSN to the FC, from the WSN to the RIS, and from the RIS to the FC, respectively.
The received signal vector $\bm{y}\in\mathbb{C}^{N}$ at the FC is:
\begin{align}
\bm{y}= & \left(\bm{G}\bm{\Theta}\bm{H}^{r}+\bm{H}^{d}\right)\bm{D}_{\alpha}^{1/2}\,\bm{x}+\bm{w}=\bm{H}^{e}(\bm{\Theta})\,\bm{D}_{\alpha}^{1/2}\,\bm{x}+\bm{w}\label{eq: signal_model}
\end{align}
where $\bm{x}\in\mathcal{X}^{K}$ and $\bm{w}\sim\mathcal{N}_{\mathbb{C}}(\bm{0}_{N},\ensuremath{\sigma_{w}^{2}\bm{I}_{N})}$ are the transmitted signal and noise vectors, respectively. 
In Eq.~\eqref{eq: signal_model}, the diagonal matrix $\bm{\Theta}=\mathrm{diag}(e^{j\varphi_{1}},\ldots,e^{j\varphi_{M}})$, $0\leq\varphi_{m}<2\pi$, $\forall m=1,\ldots M$, collects the RIS phase-shifts. Conversely, the matrix $\bm{D}_{\alpha}=\mathrm{diag}(\alpha_{1},\ldots,\alpha_{K})$,
with $\alpha_{k}\in\mathbb{R}^{+},\,\forall k \in \mathcal{K}$, accounts for unequal transmit energy.
Last, we have defined $\bm{H}^{e}(\bm{\Theta})\,\triangleq\left(\bm{G}\bm{\Theta}\bm{H}^{r}+\bm{H}^{d}\right)$ for compactness.
Similarly, we denote with $\bm{h}_{k}^{e}(\bm{\Theta})$ the $k$th column of composite channel $\bm{H}^{e}(\bm{\Theta})$.

The direct links between the sensors and the FC are modeled as the Rayleigh fading channels, i.e.
$\ensuremath{\bm{H}^{d}=\hat{\bm{H}}}^{d}\,\bm{D}_{wf}^{1/2}$,
where $\ensuremath{\hat{\bm{h}}_{k}^{d}}\sim\ensuremath{\mathcal{N}_{\mathbb{C}}(\bm{0}_{N},\bm{I}_{N})}$ and $\bm{D}_{wf}^{1/2}$ is a diagonal matrix accounting for the path loss between each sensor and the FC.
In contrast, we use a Rician fading model for the links ($a$) between the WSN and RIS and ($b$) between the RIS and FC.
Specifically, for ($a$) $\ensuremath{\bm{H}^{r}=[\bm{H}}_{\mathrm{LoS}}^{r}\,\bm{B}_{wr}+\hat{\bm{H}}^{r}\,(\bm{I}-\bm{B}_{wr}^{2})^{1/2}]\bm{D}_{wr}^{1/2}$, while for ($b$), the channel is modelled as $\bm{G}=\sqrt{d_{rf}}\,(b\,\bm{G}_{\mathrm{LoS}}+\sqrt{1-b^{2}}\,\hat{\bm{G}})$. 
Here, $\ensuremath{\bm{H}}_{\mathrm{LoS}}^{r}$ and $\bm{G}_{\mathrm{LoS}}$ represent the LoS terms for WSN-RIS and RIS-FC links%
\footnote{Herein we adopt functional forms of steering vectors $\bm{a}(\cdot)$ originating from a ``far-field'' assumption.
Accordingly, relevant terms depends on the azimuth-elevation angle-of-arrival pair at the FC.
Specifically, $\bm{h}_{k,\mathrm{LoS}}^{r}=\bm{a}_{\mathrm{upa}}(\theta_{k}^{\mathrm{AoA}},\vartheta_{k}^{\mathrm{AoA}})$
and $\bm{G}_{\mathrm{LoS}}=\bm{a}_{\mathrm{ula}}(\theta_{\mathrm{FC}}^{\mathrm{AoA}})\bm{a}_{\mathrm{upa}}^{\dagger}(\theta_{\mathrm{RIS}}^{\mathrm{AoD}},\vartheta_{\mathrm{RIS}}^{\mathrm{AoD}})$
denote the line-of-sight contributions of the aforementioned channels.
Finally, half-wavelength spaced elements are considered for both the planar RIS and the FC uniform linear receive array.}%
, respectively, 
while the diagonal matrices $\bm{B}_{wr}$ and $\bm{D}_{wr}^{1/2}$ (and the scalar terms $b$ and $\sqrt{d_{rf}}$) account for Rician factors and path loss.
Finally, $\ensuremath{\hat{\bm{h}}_{k}^{r}}\sim\ensuremath{\mathcal{N}_{\mathbb{C}}(\bm{0}_{M},\bm{I}_{M})}$ and $\hat{\bm{g}_{n}}\sim\ensuremath{\mathcal{N}_{\mathbb{C}}(\bm{0}_{N},\bm{I}_{N})}$ 
are the $k$th and $n$th column of $\hat{\bm{H}}^{r}$ and $\hat{\bm{G}}$, respectively, corresponding to the normalized NLOS (scattered) components of the aforementioned links.

\noindent
\textbf{Favorable propagation:} in the RISmMIMO context, the well-known favorable propagation conditions \emph{do not hold} for $N\gg K$~\cite{wang2020intelligent,zhi2021ergodic}. Specifically,
\begin{equation}
(\bm{H}^{e}(\bm{\Theta})^{\dagger}\bm{H}^{e}(\bm{\Theta}))\,/\,N\quad\approx\quad\underbrace{\bm{D}_{wf}+(\bm{H}^{r})^{\dagger}\,\bm{K}(\bm{\Theta})\,\bm{H}^{r}}_{\triangleq\bm{V}(\bm{\Theta})};\label{eq: favourable propagation}
\end{equation}
where
\begin{equation}
\bm{K}(\bm{\Theta})\triangleq d_{rf}\,\bm{\Theta}^{*}\,\left[(1-b^{2})\bm{I}_{M}+b^{2}\,\bm{a}_{M}\bm{a}_{M}^{\dagger}\right]\,\bm{\Theta}
\end{equation}
where we have used the short-hand notation $\bm{a}_{M}\triangleq\bm{a}_{\mathrm{upa}}(\theta_{\mathrm{RIS}}^{\mathrm{AoD}},\vartheta_{\mathrm{RIS}}^{\mathrm{AoD}})$.

The large-scale approximation of the Gram matrix $\bm{V}(\bm{\Theta})$ highlights its non-deterministic nature, as $\bm{K}(\bm{\Theta})$ depends on long-term channel parameters while $\bm{H}^{r}$ captures the instantaneous channel.
Equally important, the r.h.s. matrix of Eq.~\eqref{eq: favourable propagation} is non-diagonal, indicating a lack of decorrelation among channel vectors associated with different sensors, i.e. $\bm{h}_{k}^{e}(\bm{\Theta})^{\dagger}\bm{h}_{j}^{e}(\bm{\Theta})\neq0$,
even asymptotically.

\section{Fusion Rule Design for Large-Array FC}\label{sec_fusion_rule_RIS}
\noindent
\textbf{Optimal fusion rule:} the optimal test \cite{Kay1998} for this problem is
\begin{equation}
\left\{ \Lambda_{\mathrm{opt}}\triangleq\ln\left[\frac{p(\bm{y}|\mathcal{H}_{1})}{p(\bm{y}|\mathcal{H}_{0})}\right]\right\} \begin{array}{c}
{\scriptstyle \hat{\mathcal{H}}=\mathcal{H}_{1}}\\
\gtrless\\
{\scriptstyle \hat{\mathcal{H}}=\mathcal{H}_{0}}
\end{array}\gamma\label{eq:neyman_pearson_test}
\end{equation}
where $\hat{\mathcal{H}}$ is the estimated hypothesis, $\Lambda_{\mathrm{opt}}$ the log-likelihood ratio (LLR), and $\gamma$ the threshold chosen to maintain a fixed false-alarm rate or minimize error probability~\cite{Kay1998}. 
Using the independence\footnote{Indeed the directed triple formed by hypothesis, the transmitted-signal
vector and the received-signal vector satisfies the Markov property.}  of $\bm{y}$ from $\mathcal{H}_{i}$ given  $\bm{x}$, the LLR becomes:
\begin{gather}
\Lambda_{\mathrm{opt}}=\ln\left[\frac{\sum_{\bm{x}\in{\cal X}^{K}}p(\bm{y}|\bm{x})\Pr(\bm{x}|\mathcal{H}_{1})}{\sum_{\bm{x}\in{\cal X}^{K}}p(\bm{y}|\bm{x})\Pr(\bm{x}|\mathcal{H}_{0})}\right]\label{eq:LLR_RIS}\\
=\ln\left[\frac{\sum_{\bm{x}\in{\cal X}^{K}}\exp\left(-\frac{\bm{\|y}-\bm{H}^{e}(\bm{\Theta})\bm{D}_{\alpha}^{1/2}\bm{x}\|^{2}}{\sigma_{w}^{2}}\right)\Pr(\bm{x}|\mathcal{H}_{1})}{\sum_{\bm{x}\in{\cal X}^{K}}\exp\left(-\frac{\bm{\|y}-\bm{H}^{e}(\bm{\Theta})\bm{D}_{\alpha}^{1/2}\bm{x}\|^{2}}{\sigma_{w}^{2}}\right)\Pr(\bm{x}|\mathcal{H}_{0})}\right]\nonumber 
\end{gather}
\textbf{MRC rule:} the LLR expression can be simplified assuming perfect sensors~\cite{ciuonzo2015},
i.e. $\Pr(\bm{x}=\bm{1}_{K}|\mathcal{H}_{1})=\Pr(\bm{x}=-\bm{1}_{K}|\mathcal{H}_{0})=1$.
In this case $\bm{x}\in\{\bm{1}_{K},-\bm{1}_{K}\}$ and Eq.~(\ref{eq:LLR_RIS})
reduces to:
\begin{align}
\ln\left[\frac{\exp\left(-\frac{\|\bm{y}-\bm{H}^{e}(\bm{\Theta})\,\bm{D}_{\alpha}^{1/2}\,\bm{1}_{K}\|^{2}}{\sigma_{e}^{2}}\right)}{\exp\left(-\frac{\|\bm{y}+\bm{H}^{e}(\bm{\Theta})\,\bm{D}_{\alpha}^{1/2}\,\bm{1}_{K}\|^{2}}{\sigma_{e}^{2}}\right)}\right] & \propto\Re\left(\bm{a}_{\mathrm{{\scriptscriptstyle MRC}}}^{\dagger}\bm{y}\right)\triangleq\Lambda_{{\scriptscriptstyle \mathrm{MRC}}}\label{eq:MRC}
\end{align}
where $\bm{a}_{{\scriptscriptstyle \mathrm{MRC}}}\triangleq(\bm{H}^{e}(\bm{\Theta})\,\bm{D}_{\alpha}^{1/2}\,\bm{1}_{K})$
and terms independent from $\bm{y}$ have been incorporated in $\gamma$
as in Eq.~(\ref{eq:neyman_pearson_test}). 
It is important to note that MRC is \emph{sub-optimal} since sensor local decisions are rarely perfect. 
However, \cite{Ciuonzo2012} demonstrated that MRC approximates the optimal test in Eq.~(\ref{eq:LLR_RIS}) under low SNR when sensors' local performances and path losses are identical.
To address this limitation, \cite{ciuonzo2015} proposed a modified MRC that includes diagonal scaling of the matched-filtered data $\bm{H}^{e}(\bm{\Theta})^{\dagger}\bm{y}$ to correct the large-scale approximation of the Gram matrix $(\bm{H}_{d}^{\dagger}\bm{H}_{d})\,/\,N\approx\bm{D}_{wf}$, thereby enhancing the benefits of favorable propagation and providing a more accurate counting rule.

In RISmMIMO, the favorable propagation condition does not hold (see Eq.~\eqref{eq: favourable propagation}). Hence, for large $N$, MRC statistic approximates
\begin{equation}
(\Lambda_{\mathrm{{\scriptscriptstyle MRC}}}/N)\approx\Re\{\bm{1}_{K}^{T}\bm{D}_{\alpha}^{1/2}\,\bm{V}(\bm{\Theta})\,\bm{D}_{\alpha}^{1/2}\,\bm{x}\}+\Re\{w{}_{\mathrm{mrc}}\}\label{eq: asymptotic_form_MRC}
\end{equation}
where $w_{\mathrm{mrc}}\sim\mathcal{N}_{\mathbb{C}}(0,\frac{\sigma_{w}^{2}}{N}\bm{1}_{K}^{T}\bm{D}_{\alpha}^{1/2}\,\bm{V}(\bm{\Theta})\,\bm{D}_{\alpha}^{1/2}\,\bm{1}_{K})$.

\noindent
\textbf{Modified MRC-1/2}: To develop an analogous modified MRC fusion rule (m\textsc{MRC-1}) for RISmMIMO, we define
\begin{equation}
\bm{a}_{\mathrm{m{\scriptscriptstyle MRC}-1}}\triangleq(\bm{H}^{e}(\bm{\Theta})\,\bm{V}^{-1}(\bm{\Theta})\,\bm{D}_{\alpha}^{-1/2}\bm{1}_{K})\,,
\end{equation}
which yields (for $N \gg K$):
\begin{equation}
(\Lambda_{\mathrm{m{\scriptscriptstyle MRC-1}}}/N)\approx\sum_{k=1}^{K}x_{k}+\Re\{w_{\mathrm{mMRC-1}}\}\label{eq: asymptotic_form_mMRC}
\end{equation}
where $w_{\mathrm{mMRC-1}}\sim\mathcal{N}_{\mathbb{C}}(0,\frac{\sigma_{w}^{2}}{N}\bm{1}_{K}^{T}\bm{D}_{\alpha}^{-1/2}\,\bm{V}(\bm{\Theta})^{-1}\,\bm{D}_{\alpha}^{-1/2}\,\bm{1}_{K})$.
Despite its apparent advantages, this rule requires $\bm{V}(\bm{\Theta})^{-1}$, thus necessitating knowledge of $\bm{H}^{r}$ (see Eq.~\eqref{eq: favourable propagation}). Consequently, estimating all relevant channel terms requires $(M+2K-1)$ pilot symbols~\cite{wang2020channel}, compared to the usual $K$ symbols needed for the standard MRC fusion rule (which requires $\bm{H}^{e}(\bm{\Theta})$).
To address this, we also explore a second version of the modified MRC fusion rule, denoted m\textsc{MRC-2}, where $\bm{V}(\bm{\Theta})$ is replaced by its expectation $\bar{\bm{V}}(\bm{\Theta})$, namely
\begin{equation}
\bm{a}_{\mathrm{m{\scriptscriptstyle MRC-2}}}\triangleq\left(\bm{H}^{e}(\bm{\Theta})\,\bar{\bm{V}}(\bm{\Theta})^{-1}\,\bm{D}_{\alpha}^{-1/2}\bm{1}_{K}\right)
\end{equation}
The explicit form of $\bar{\bm{V}}(\bm{\Theta})$ can be shown to be equal (proof is not reported for brevity):
\begin{gather}
\bar{\bm{V}}(\bm{\Theta})=\bm{D}_{wf}+\bm{D}_{wr}\,(\bm{I}_{K}-\bm{B}_{wr}^{2})\,+\nonumber \\
(\ensuremath{\bm{H}}_{\mathrm{LoS}}^{r}\,\bm{B}_{wr}\,\bm{D}_{wr}^{1/2})^{\dagger}\,\bm{K}(\bm{\Theta})\,(\ensuremath{\bm{H}}_{\mathrm{LoS}}^{r}\,\bm{B}_{wr}\,\bm{D}_{wr}^{1/2})
\end{gather}
\textbf{Zero Forcing Combiner}: building on recent successes with  zero-forcing techniques~\cite{zhi2021ergodic,zhi2022ris} instead of matched filters, we introduce a new fusion rule using a zero-forcing combiner. Specifically, we define $\Lambda_{{\scriptscriptstyle \mathrm{ZFC}}}\triangleq\Re(\bm{a}_{\mathrm{{\scriptscriptstyle ZFC}}}^{\dagger}\,\bm{y})$,
where
\begin{equation}
\bm{a}_{\mathrm{{\scriptscriptstyle ZFC}}}\triangleq\ensuremath{(\bm{H}^{e}(\bm{\Theta})\,(\bm{H}^{e}(\bm{\Theta})^{\dagger}\bm{H}^{e}(\bm{\Theta}))^{-1}\bm{D}_{\alpha}^{-1/2}\,\bm{1}_{K})}
\end{equation}
In such a case, the fusion statistic can be rewritten ($\forall N \geq K$) as
\begin{equation}
(\Lambda_{\mathrm{ZFC}}/N)=\sum_{k=1}^{K}x_{k}+\Re\{w_{\mathrm{ZFC}}\}\label{eq: asymptotic_form_ZFC}
\end{equation}
where
\begin{equation}
w_{\mathrm{ZFC}}\sim\mathcal{N}_{\mathbb{C}}(0,\frac{\sigma_{w}^{2}}{N}\bm{1}_{K}^{T}\bm{D}_{\alpha}^{-1/2}\left(\frac{\bm{H}^{e}(\bm{\Theta})^{\dagger}\bm{H}^{e}(\bm{\Theta})}{N}\right)^{-1}\bm{D}_{\alpha}^{-1/2}\bm{1}_{K})
\end{equation}
It can be observed that the ZFC rule approximately matches the distribution of m\textsc{MRC-1} as $N$ becomes large. 
Table~\ref{tab:fusion_rules_comparison}  summarizes ($i$) the complexity associated with each decision vector $\bm{x}$ and each new channel realization, and ($ii$)  the CSI requirements for the three proposed fusion rules, comparing them with MRC and LLR.

\begin{table}[t]
\centering
\caption{Comparison for the fusion rules considered.}
\label{tab:fusion_rules_comparison}
\resizebox{\columnwidth}{!}{ 
\begin{threeparttable}
\begin{tabular}{l c c c}
\toprule
\textbf{Fusion} & \textbf{Complexity}  & \textbf{CSI Reqs.} \\
\textbf{Rule} & \textbf{(new $\bm{x}$ / channel)}  & \textbf{(pilots)} \\
\midrule
\rowcolor{blue!15}
\textsc{LLR \cite{mudkey2022wireless}} & $\mathcal{O}(N 2^K)$  /   $\mathcal{O}(N K 2^K)$  & $K$   \\ 
\rowcolor{red!35}
\textsc{MRC \cite{mudkey2022wireless}} & $\mathcal{O}(N)$ /   $\mathcal{O}(NK)$ & $K$ \\
\rowcolor{modMRCcolor!60}
m\textsc{MRC-1} & $\mathcal{O}(N)$ /   $\mathcal{O}(M^2 K + K^2 M + K^3)$  & $M+2K-1$  \\
\rowcolor{modMRCcolor2!40}
m\textsc{MRC-2} & $\mathcal{O}(N)$ /   $\mathcal{O}(N K^2)$ & $K$  \\ 
\rowcolor{ZFCcolor!20}
\textsc{ZFC} & $\mathcal{O}(N)$  /   $\mathcal{O}(N K^2 + K^3)$ & $K$  \\ 
\bottomrule
\end{tabular}
\end{threeparttable}
}
\end{table}

\section{RIS design via long-term channel statistics}\label{sec_RIS_design_LT}
\textbf{Definition of RIS design objective:} by looking at large-array expressions in \eqref{eq: asymptotic_form_mMRC} and \eqref{eq: asymptotic_form_ZFC}, we observe that both rules have the form of noisy counting rules, i.e. $\approx \sum_{k=1}^{K}x_{k}+\Re\{w\}$.
Accordingly, the RIS design can be devoted to minimize the variance of the associated noise term. 
This is tantamount to minimizing the term $\frac{\sigma_{w}^{2}}{N}\bm{1}_{K}^{T}\bm{D}_{\alpha}^{-1}\,\bm{V}(\bm{\Theta})^{-1}\,\bm{D}_{\alpha}^{-1}\,\bm{1}_{K}$, since the variance of $w_{\mathrm{ZFC}}$ also converges to the same value when $N \gg K$.
Furthermore, to avoid RIS design based on instantaneous CSI (because of $\bm{V}(\bm{\Theta})$), we consider 
\emph{dominant LoS conditions} (only at the design phase) for both the WSN-RIS and RIS-FC links, namely $\bm{B}_{wr}\approx\bm{I}_{K}$ and $b\approx1$.
When these assumptions hold, the matrix $\bm{V}(\bm{\Theta})$ converges to its expected form $\bar{\bm{V}}(\bm{\Theta})$, and both share the simplified expression: 
\begin{equation}
\ensuremath{\bm{V}_{\mathrm{LoS}}(\bm{\Theta})\triangleq\bm{D}_{wf}}+d_{rf}(\bm{H}_{\mathrm{LoS}}^{r}\,\bm{D}_{wr}^{1/2})^{\dagger}\,\bm{\Theta}^{*}\bm{a}_{M}\bm{a}_{M}^{\dagger}\bm{\Theta}\,(\bm{H}_{\mathrm{LoS}}^{r}\,\bm{D}_{wr}^{1/2})
\end{equation}
The above matrix \emph{only depends on (long-term) channel statistics}.
As a result, the formulated optimization problem is the following:
\begin{equation}
\mathcal{P}_{1}\,:\begin{array}{c}
\underset{\bm{\Theta}}{\mathrm{minimize}}\quad\bm{1}_{K}^{T}\bm{D}_{\alpha}^{-1/2}\,\bm{V}_{\mathrm{LoS}}^{-1}(\bm{\Theta})\,\bm{D}_{\alpha}^{-1/2}\,\bm{1}_{K}\qquad\\
\mathrm{subject\,to\quad\bm{\Theta}=\mathrm{diag}}(e^{j\varphi_{1}},\ldots,e^{j\varphi_{M}})
\end{array}
\end{equation}
By ($i$) exploiting the Sherman–Morrison formula~\cite{bernstein2009matrix}, ($ii$) neglecting terms independent on $\bm{\Theta}$, and ($iii$) re-arranging the obtained objective (e.g. also expressing it in terms of $\bm{\theta}\triangleq\begin{bmatrix}e^{j\varphi_{1}} & \cdots & e^{j\varphi_{M}}\end{bmatrix}^{T}$), problem $\mathcal{P}_1$ can be shown to be equivalent to:
\begin{equation}
\mathcal{P}_{2}\,:\begin{array}{c}
\underset{\bm{\theta}}{\mathrm{maximize}}\quad g(\bm{\theta})=\bm{\theta}^{T}\bm{v}_{1}\bm{v}_{1}^{\dagger}\bm{\theta^{*}}\,/\,(\bm{\theta}^{T}\,\bm{\Xi}\,\bm{\theta}^{*})\\
\mathrm{subject\,to\quad}|\theta_{m}|=1,\hfill m=1,\ldots,M
\end{array}
\end{equation}
where $\bm{v}_{1}\triangleq\bm{S}_{1}\,\bm{D}_{wf}^{-1}\,\bm{D}_{\alpha}^{-1/2}\,\bm{1}_{K}$ and
$\,\bm{\Xi}\triangleq\left((1/M)\,\bm{I}_{M}+\bm{S}_{1}\bm{D}_{wf}^{-1}\,\bm{S}_{1}^{\dagger}\right)$. In the latter two terms the auxiliary matrix definition
\begin{equation}
\bm{S}_{1}\triangleq\sqrt{d_{rf}}\,\mathrm{diag}(\bm{a}_{M})^{*}\,\bm{H}_{\mathrm{LoS}}^{r}\,\bm{D}_{wr}^{1/2}
\end{equation}
has been exploited.
The problem $\mathcal{P}_2$ can be solved by resorting to the well-known MM framework~\cite{sun2016majorization}, as described hereinafter.

\noindent
\textbf{MM-based optimization strategy:} 
assuming the value of $\bm{\theta}$ in the $\ell$th iteration is denoted as $\bm{\theta}_{(\ell)}$, MM aims to construct a lower bound on the objective function $g(\bm{\theta})$ that touches the objective function at point $\bm{\theta}$, denoted as $f(\bm{\theta}|\bm{\theta}_{(\ell)})$.
We adopt this lower bound as a surrogate objective function, and the maximizer of this surrogate objective function is then taken as the value of $\bm{\theta}$ in the next iteration, i.e., $\bm{\theta}_{(\ell+1)}$. 
Consequently, $g(\bm{\theta})$ increases monotonically, i.e. $g(\bm{\theta}_{(\ell+1)})\geq g(\bm{\theta}_{(\ell)})$, achieving first-order optimality.
Success hinges on finding a surrogate $f(\bm{\theta}|\bm{\theta}_{(\ell)})$ where the maximizer $\bm{\theta}_{(\ell+1)}$ is easy to compute. 
For problem  $\mathcal{P}_2$, a suitable surrogate of $g(\bm{\theta})$ is the following (leveraging convexity of $\bm{\theta}^{T}\bm{v}_{1}\bm{v}_{1}^{\dagger}\bm{\theta^{*}}\,/\,c$ in $\{\bm{\theta},c\}$):
\begin{gather}
f(\bm{\theta}|\bm{\theta}_{(\ell)})=2\frac{\Re\{\bm{\theta}_{(\ell)}^{\dagger}\,\bm{v}_{1}\bm{v}_{1}^{\dagger}\,\bm{\theta}\}}{\bm{\theta}_{(\ell)}^{\dagger}\,\bm{\Xi}\,\bm{\theta}}-\frac{\bm{\theta}_{(\ell)}^{\dagger}\bm{v}_{1}\bm{v}_{1}^{\dagger}\,\bm{\theta}_{(\ell)}^{\dagger}}{(\bm{\theta}_{(\ell)}^{\dagger}\,\bm{\Xi}\,\bm{\theta}_{(\ell)})^{2}}+\mathrm{const}
\end{gather}
Accordingly, the MM problem to be solved is the following
$\mathcal{P}_{\mathrm{{\scriptscriptstyle MM}}}\,:\,\bm{\theta}_{(\ell+1)}=\underset{|\theta_{m}|=1}{\arg\max}\,f(\bm{\theta}|\bm{\theta}_{(\ell)})$.
Remarkably, the optimal solution of $\mathcal{P}_{\mathrm{{\scriptscriptstyle MM}}}$ is in \emph{closed-form}.
Hence, problem $\mathcal{P}_2$ is solved by \emph{iteratively} updating the RIS phase vector as
\begin{gather}
\angle\bm{\theta}_{(\ell+1)}=\angle\left(\frac{\bm{v}_{1}\bm{v}_{1}^{\dagger}\,\bm{\theta}_{(\ell)}}{\left(\bm{\theta}_{(\ell)}\right)^{\dagger}\bm{\Xi}\,\bm{\theta}_{(\ell)}}-\right.\label{eq: RIS optimization (Step 2) fc}\\
\left.\frac{\left(\bm{\theta}_{(\ell)}\right)^{\dagger}\,\bm{v}_{1}\bm{v}_{1}^{\dagger}\,\bm{\theta}_{(\ell)}}{\left(\left(\bm{\theta}_{(\ell)}\right)^{\dagger}\bm{\Xi}\,\bm{\theta}_{(\ell)}\right)^{2}}\,\left(\bm{\Xi}-\lambda_{max}(\bm{\Xi})\,\bm{I}_{M}\right)\,\bm{\theta}_{(\ell)}\right)\nonumber 
\end{gather}
where $\lambda_{max}(\bm{\Xi})$ is the highest eigenvalue of $\bm{\Xi}$.
The procedure is triggered by setting randomly the RIS phases in $\bm{\theta}_{(0)}$.
We remark that the designed vector $\bm{\theta}^{\star}$ depends only on long-term channel statistics. Thus, it should be executed once every large number of time slots.

\section{Simulation Results}\label{sec:sim_results}
\textbf{Simulation setup:} we consider a WSN made of $K=10$ sensors whose decisions on the phenomenon of interest are conditionally independent and identically distributed (i.i.d.), i.e.,
$P(\bm{x}|\mathcal{H}_{i})=\prod_{k=1}^{K}P(x_{k}|\mathcal{H}_{i})$, $(P_{D,k},P_{F,k})\triangleq(0.5,0.05)$, $k\in\mathcal{K}$, similarly to ~\cite{Chen2004}.
The locations of the WSN, RIS and FC are as follows: the sensors are uniformly distributed at random in the square $[0,40]\times[0,40]\,\mathrm{m}^{2}$ placed on the ground. Conversely, the RIS and the FC are located at $[40, 20, 5]\,\mathrm{m}$ and $[65, 40, 2]\,\mathrm{m}$, respectively.
In what follows we consider a RIS with $M=25$ elements.
For simplicity, the sensors are assumed to have equal transmit energy, namely $\bm{D}_{\alpha}=\bm{I}_K$.
The path loss model considered is $P(d,\nu)=\mu\,(d/d_{0})^{-\nu}$, where $\mu = - 20\,\mathrm{dB}$ is the path loss attenuation at the reference distance of $d_0=1$ m, while $\nu$ denotes the path loss exponent. The latter is set to $\nu_1=2$ for both WSN-to-RIS and RIS-to-FC links, and $\nu_2=4$ for WSN-FC (obstructed) links.
When not otherwise stated, the Rician factors are generated randomly $\in (10,20)\,\unit{dB}$.
Finally, the noise variance $\sigma_{w}^{2}$ is set to $-70$ dBm.

\noindent
\textbf{Performance metrics and upper-bounds:} we analyze the performance of the fusion rules in terms of the probabilities of \textit{false alarm} $P_{F_{0}}\triangleq\Pr\{\Lambda>\gamma|H_{0}\}$ and \textit{detection} $P_{D_{0}}\triangleq\Pr\{\Lambda>\gamma|H_{1}\}$.
To benchmark degradation of detection performance due to the interfering distributed MIMO channel (and the benefits arising from the combined use of a  RIS and a massive array at the FC), we also report the ``observation bound'', i.e. the performance of the optimal decision fusion rule in an \textbf{ideal} channel condition, given by $P_{D_{0}}^{\text{ob}}  =\sum_{i=\nu}^{K}\binom{K}{i}
(P_{D})^{i} \, (1-P_{D})^{K-i}$ and
$P_{F_{0}}^{\text{ob}} =\sum_{i=\nu}^{K}\binom{K}{i}
(P_{F})^{i} \, (1-P_{F})^{K-i}$,
where $\nu\in\{0,\ldots K\}$ is a discrete threshold.

\begin{figure}
\centering{}\includegraphics[width=0.86\columnwidth]{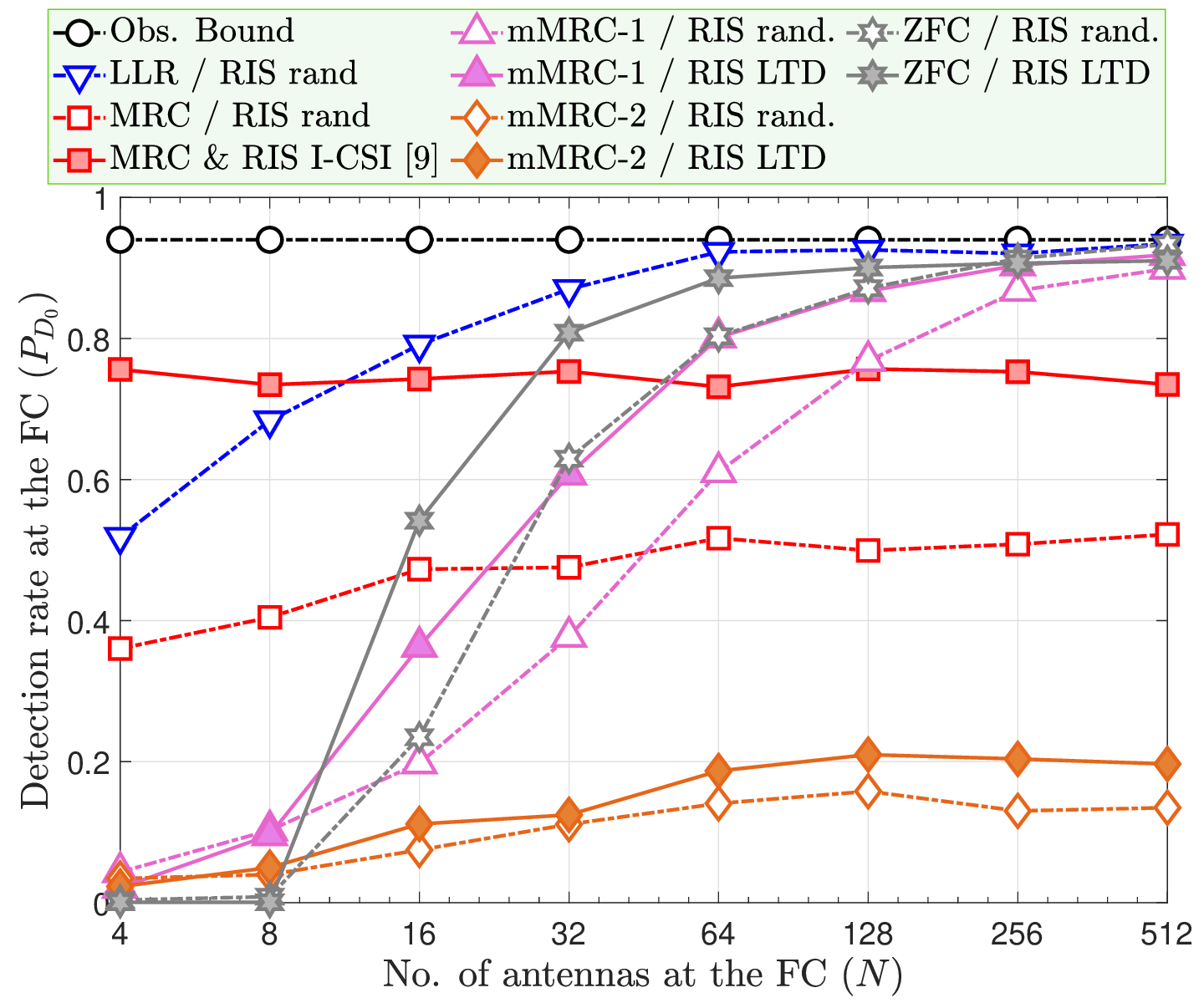}\caption{$P_{D_{0}}$ vs $N$ of the considered rule/RIS configurations.  Setup: FC false-alarm rate is set to $P_{F_{0}}=0.01$;  RIS with $M=25$ elements.\label{fig:PD0_vs_N}}
\end{figure}

\noindent
\textbf{Assessing large-array benefits:} Fig.~\ref{fig:PD0_vs_N} reports the $P_{D_{0}}$ versus the number of receive antennas $N$ at the FC.
The aim is to understand the interplay between the number of receive antennas and the RIS benefit for a relevant number of atoms.
The LLR shows clear improvement in detection rate with increasing $N$, while MRC performance remains constant, as predicted by \eqref{eq: asymptotic_form_MRC}.
This trend persists even when MRC and RIS are jointly optimized using instantaneous CSI~\cite{mudkey2022wireless} (curve \say{MRC \& RIS I-CSI}). 
In contrast, m\textsc{MRC-1} and \textsc{ZFC} significantly enhance detection performance with larger $N$, with further gains from the proposed long-term RIS design (curves \say{RIS LTD}). 
For example, when $N\approx 32$ (or $N\approx64$), \textsc{ZFC} (or m\textsc{MRC-1}) surpasses the joint MRC-RIS design with instantaneous CSI.
However, m\textsc{MRC-2}, despite benefiting from the RIS design, shows poor detection performance, even worse than MRC. This issue is explored further in the subsequent analysis.

\begin{figure}
\centering{}\includegraphics[width=0.86\columnwidth]{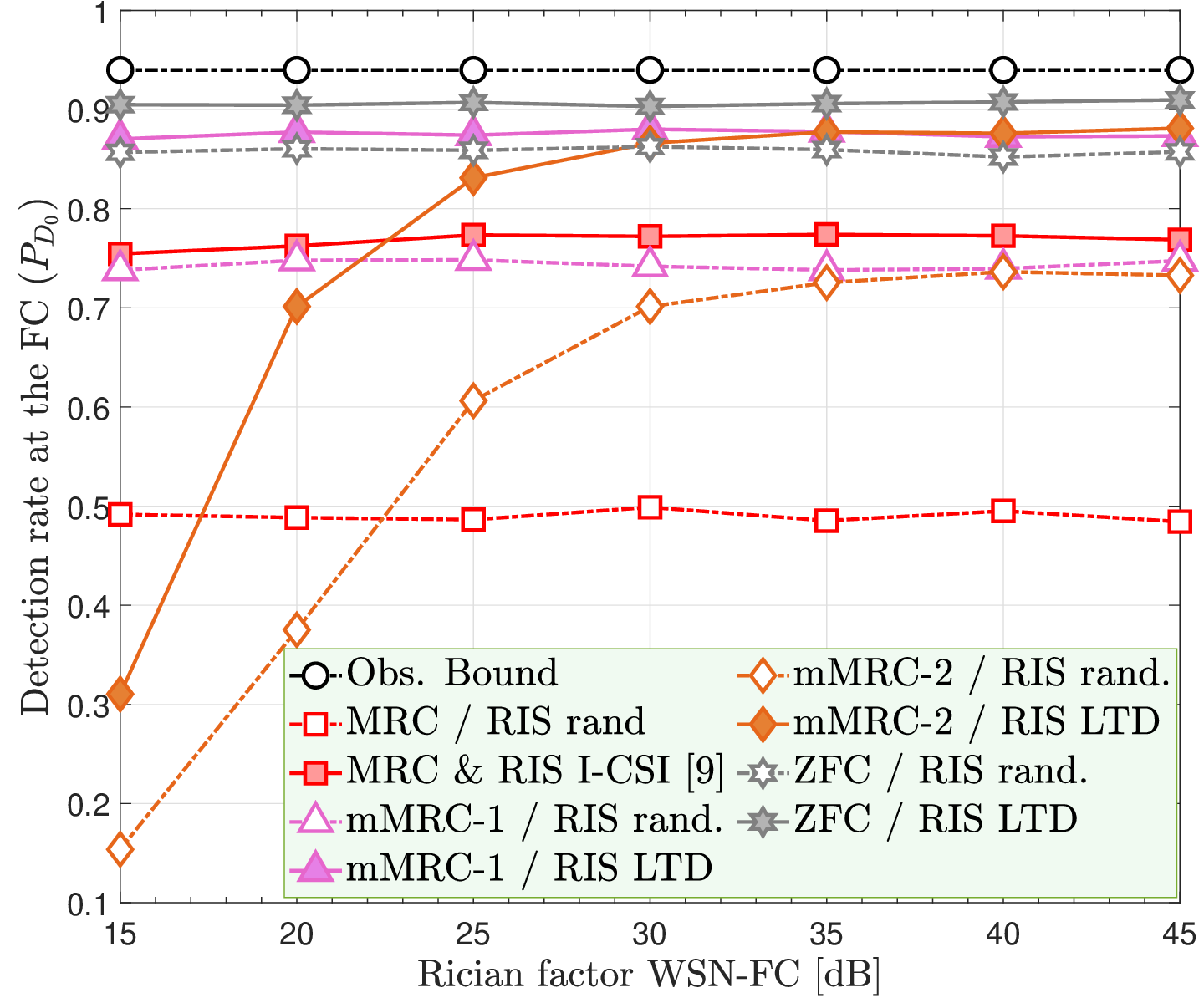}\caption{$P_{D_{0}}$ vs WSN-FC Rician factor [dB] for the considered rule/RIS configurations.  Setup: FC false-alarm rate is set to $P_{F_{0}}=0.01$;  RIS with $M=25$ elements and $N=128$ antennas at the FC.\label{fig:PD0_vs_Rician}}
\end{figure}

\noindent
\textbf{Assessing sensitivity to LoS conditions:} to investigate why m\textsc{MRC-2} performs worse m\textsc{MRC-1} and \textsc{ZFC}, Fig.~\ref{fig:PD0_vs_Rician} varies the Rician factor as $\bm{B}_{wr} = b_{wr} \bm{I}_K$ within $(15,45)\,\unit{dB}$ while fixing $b$ at $20\,\unit{dB}$, with $N=128$ at the FC. 
The aim is to examine how detection rates for these fusion rules depend on the LoS conditions between the WSN and RIS. The figure reveals that m\textsc{MRC-2} performance improves significantly with stronger LoS components from the WSN to the FC, with higher Rician factors leading to better detection in both optimized and non-optimized scenarios. 
As the Rician factor increases, m\textsc{MRC-2} surpasses the joint MRC+RIS design with full CSI, approaching the performance of m\textsc{MRC-1} (noting that $\bm{V}(\bm{\Theta})\rightarrow\bar{\bm{V}}(\bm{\Theta})$ as $\bm{B}_{wr}\rightarrow\bm{I}_{K}$). 
In contrast, m\textsc{MRC-1} and \textsc{ZFC} maintain consistently high performance, similar to the joint MRC+RIS design, due to their use of the true matrix $\bm{H}^r$.

\section{Conclusions and Future Directions}
\label{Conclusions}

This work represents an initial effort to combine RIS with large arrays for channel-aware decision fusion in a virtual MIMO setup. 
To address the high complexity and information requirements of LLR, we introduced three sub-optimal linear fusion rules to exploit the benefits of large arrays, contrasting with standard alternatives like MRC. 
Additionally, we developed a RIS design based on long-term channel statistics, leveraging dominant LoS conditions. This approach led to a straightforward iterative optimization procedure within the MM framework. 
Simulation results highlight the effectiveness of integrating RIS with large arrays at the FC for enhancing WSN decision fusion, even with sub-optimal designs. 
\emph{Future work} will explore: ($i$) similar fusion rules and RIS designs for large-RIS or large-FC+RIS scenarios; ($ii$) theoretical performance analysis in these settings; ($iii$) the use of multiple RISs;  ($iv$) design based on practical RIS codebooks; and ($v$) handling of imperfect CSI.

\bibliographystyle{IEEEbib}
\bibliography{refs}

\end{document}